\documentclass{article}
\usepackage[acronym]{glossaries-extra}
\usepackage{xcolor}
\usepackage{tikz}
\usepackage{listings}
\usepackage{amsmath}
\usepackage{algorithm}
\usepackage{graphicx}
\usepackage[utf8]{inputenc}
\usepackage[english]{babel}
\usepackage{authblk}
\usepackage{algpseudocode}
\usepackage{listings}
\usepackage{subcaption}
\usepackage{siunitx}
\usepackage[colorlinks=true, allcolors=blue]{hyperref}
\usepackage[capitalise]{cleveref}
\usepackage{float}
\usepackage[letterpaper,top=2cm,bottom=2cm,left=3cm,right=3cm,marginparwidth=1.75cm]{geometry}
\usepackage[style=numeric,sorting=none]{biblatex}
\setabbreviationstyle[acronym]{long-short}
\newacronym{llm}{LLM}{Large Language Model}
\newacronym{mc}{MC}{Monte Carlo}
\newacronym{mcp}{MCP}{Model Context Protocol}
\newacronym{ks}{KS}{Kolmogorov-Smirnov}
\newacronym{topas}{TOPAS}{TOol for PArticle Simulation}
\usetikzlibrary{shapes.geometric, shapes.misc, arrows.meta, positioning, calc}
\glsdisablehyper
\title{Generating a simple silicon detector simulation with an open-weight large language model}
\author[1]{Charlie H R Bamber}
\author[1]{Ming-Liang Wong\thanks{Corresponding author. Email: \texttt{mlwong@liverpool.ac.uk} } }
\affil[1]{Department of Physics, University of Liverpool, Oliver Lodge Building, Oxford St., Liverpool L69 7ZE, UK}
\date{}

\begin{document}
\maketitle
\begin{abstract}
Recent \gls{llm} developments have seen AI-generated code become commonplace in the software development industry. Without tool access, \gls{llm}s cannot execute simulations, and their knowledge of toolkits such as Geant4 is limited. We demonstrate here, a detector simulation workflow with a Qwen3.6-27B model with Q4\_K\_M GGUF quantisation together with custom resources and tools served via \gls{mcp} servers to generate a particle detector simulation. In 50 independent runs, 42(29--56)\% at 95\% CI produced energy-deposit distributions statistically consistent with a reference simulation, measured against a reference simulation. These results provide a proof of concept that \gls{mc} simulations can be generated using text-based prompts with an \gls{llm}, given the necessary resources and tools. More extensive resources, such as example code and additional documentation, are likely to be required for more complex simulation tasks.
\end{abstract}
\section{Introduction}
\gls{llm}s are increasingly being used for automating time-consuming and labour-intensive software tasks, such as code generation. \gls{llm} agents have been used to automate parts of particle physics data analysis~\cite{GendreauDistler:2025hep, Moreno:2026hep, Diefenbacher:2025agents}. However, their use for directly generating code and running \gls{mc} simulations of physics detectors, which are central to particle physics, has received less attention~\cite{Chung:2026agentic, Zoccheddu_2026}. Such simulations are part of the detector design and analysis steps of almost all experiments to understand the event features of signal processes and their backgrounds. This is crucial for relating experimentally measured variables to the parameters of the underlying theory. \gls{mc} simulations require expertise in both software frameworks and particle physics. Moreover, detector design changes take significant time to incorporate into the simulation and require further software debugging and physics validation. In this paper, we demonstrate the potential use and motivate future studies of \gls{llm}s to generate particle detector \gls{mc} simulations from text prompts when given access to tools and resources.

\gls{llm}-based assistants have been shown to increase developer productivity in field experiments~\cite{Cui_Demirer_Jaffe_Musolff_Peng_Salz_2026}. Applied to \gls{mc} simulation, an \gls{llm} that generates, runs and checks simulation code could offer several benefits. First, it lowers the barrier of entry. A detector configuration described in plain text can be translated into a working simulation without detailed knowledge of the toolkit's syntax, allowing physicists without simulation expertise explore design options. Second, it shortens the iteration cycle. Changes to geometry, materials or beam conditions can be requested in a sentence rather than implemented and debugged by hand. Third, the generated configuration files remain human-readable. So every simulation can be inspected, version-controlled and rerun. Finally, coupling the model to automated checks, such as syntax validation and statistical comparison with reference distributions, catches many errors before a person reviews the output. Running an open-weight model locally adds further advantages. Proprietary detector designs never leave the institution, there is no per-token cost, and the exact model version can be fixed so that results stay reproducible. These benefits depend on the generated simulations being correct, so physics validation by an expert remains essential.

\section{Methodology}
We construct a framework that turns a user prompt into a histogram of a simulated observable, a step common to many detector studies. The user-facing interface can be managed by any messaging service with endpoints that supports bot\footnote{The messaging service used is Mattermost 11.11.0.} accounts. The \gls{mc} simulations are created using \gls{topas}~\cite{TOPAS, Faddegon:2020topas}, a Geant4 simulation toolkit~\cite{Agostinelli:2002hh, Allison:2006ve, Allison:2016lfl} wrapper with a text-based parameter control system that allows users to assemble and control simulation objects, scorers, and particle sources. The text-based nature of the \gls{topas} parameter files is well suited to \gls{llm} text generation, since it usually requires no Geant4 C++ code.

Qwen3.6-27B~\cite{qwen3techreport} is chosen as the open-weight model. It initialises with a system prompt preloading instructions as detailed in Appendix~\ref{app:systemprompt}. An \gls{mcp}~\cite{modelcontextprotocolSpecificationModel} server and the messaging service bot are started. Once the server and scripts are running, the messaging service bot is ready to accept user prompts. The bot receives a prompt sent via the messaging service, and Qwen3.6-27B decides iteratively if it requires tools from the \gls{mcp} server based on the prompt. These include the necessary functions for creating \gls{topas} files, syntax checking, running \gls{topas} simulations, reading ROOT output files, running \gls{ks} tests, and uploading images to the messaging service. The results of the tools are returned to the model at each iteration. The final response is posted by the bot on the messaging system.

\begin{figure}[H]
    \centering
    \includegraphics[width=0.5\linewidth]{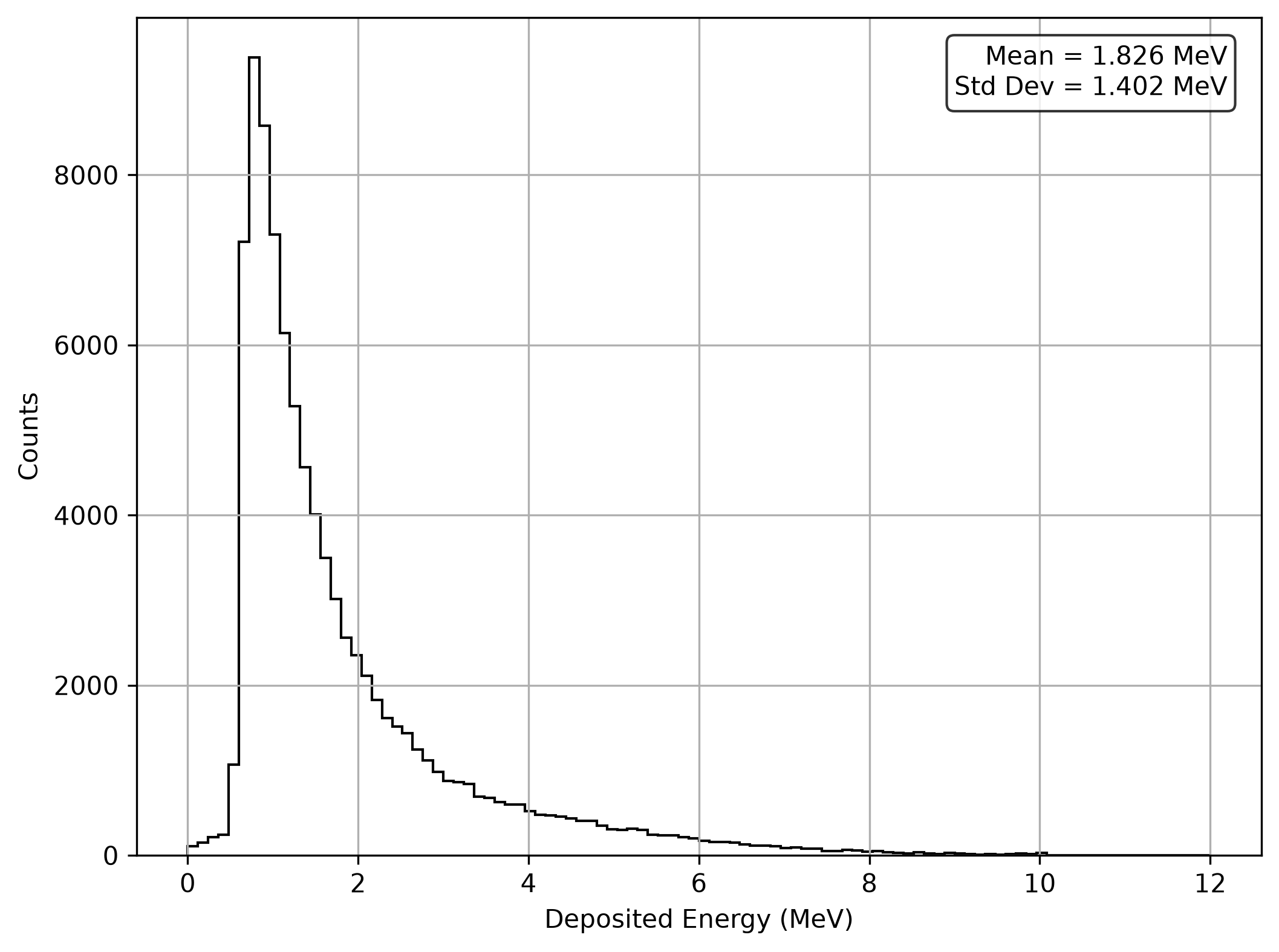}
    \caption{Energy deposit of a \SI{10}{MeV} electron beam on \SI{2}{mm} thick silicon. $10^5$ histories generated from a Gaussian beam with \SI{0.01}{cm} spread.}
    \label{fig:electron_silicon_energy_deposit}
\end{figure}

A generated reference simulation of an electron beam on silicon (\Cref{fig:electron_silicon_energy_deposit}) is chosen for its simplicity as the test case for debugging and development. From the initial user prompt shown in Appendix~\ref{app:prompt1}, a \gls{topas} parameter file is generated on the server. The syntax of this parameter file is checked using tools available to the \gls{llm} via an \gls{mcp} server. 

\subsection{Large Language Model} 
Ollama\footnote{Ollama 0.30.7}~\cite{ollama_documentation} is an open-source software platform that allows the download, management, and running of \gls{llm}s locally rather than relying on external cloud servers. It is run on an NVIDIA A100 80GB PCIe GPU~\cite{nvidiaA100}. This allows control over model parameters, system prompts, tool calling, model selection, and hardware resources. The Qwen3.6-27B Q4\_K\_M GGUF quantisation model with thinking mode enabled was used both to develop the resources and tools and for all evaluation runs. It is further configured to generate and run \gls{mc} simulations by using a model file to define the sampling parameters and the system prompt (Appendix~\ref{app:systemprompt}). The user interacts with the \gls{llm} through the messaging service. Outputs such as plain text, images in the form of histograms and graphs, and other file types, including text, Python, and C++ files, can be sent during the \gls{llm}'s processing loop using dedicated \gls{mcp} tools. 

\subsection{Tools and Resources}

\begin{figure}[H]
    \centering
    \begin{tikzpicture}[
        font=\sffamily,
        base/.style     = {draw=black!65, line width=0.9pt, align=center},
        process/.style  = {base, rectangle, rounded corners=6pt,
                           text width=2.5cm, minimum height=1.0cm},
        terminal/.style = {base, rounded rectangle, rounded rectangle arc length=180,
                           text width=2.2cm, minimum height=0.7cm},
        decision/.style = {base, diamond, aspect=1.7, inner sep=1pt,
                           text width=1.8cm, minimum height=1.0cm},
        arrow/.style    = {draw=black!65, line width=0.9pt,
                           -{Stealth[length=2mm,width=1.8mm]}},
        lbl/.style      = {font=\sffamily\bfseries\small, inner sep=3pt}
      ]
      % ---- geometry -------------------------------------------------------
      \def\colA{0}      % left column
      \def\colB{5}      % middle column
      \def\colC{10}     % right column
      \def\loopx{-2.7}  % x-position of the left-hand feedback bus

      \node[terminal] (request)  at (\colA, 0)      {Request\\Received};
     
      \node[process]  (evaluate) at (\colA,-2)      {\gls{llm} Evaluation};
      \node[process]  (gencode)  at (\colB,-2)      {\gls{llm}\\Generates Code};
     
      \node[decision] (syntaxq)  at (\colA,-4.6)    {Syntax\\Errors?};
      \node[process]  (checker)  at (\colB,-4.6)    {\gls{topas} Syntax\\Checker};
     
      \node[process]  (runtopas) at (\colA,-7.2)   {Run \gls{topas}};
      \node[process]  (analyse)  at (\colB,-7.2)   {Kolmogorov-\\Smirnov Test};
      \node[process]  (rootobj)  at (\colC,-9.5)   {Get ROOT\\Objects};
     
      \node[decision] (topasq)   at (\colA,-9.5)   {Runtime\\Error?};
      \node[decision] (acceptq)  at (\colB,-9.5)   {$p_{KS} > 0.01$};
      \node[process]  (summary)  at (\colB,-11.5)   {\gls{llm} summary};
      \node[process]  (upload)   at (\colC,-11.5)   {Upload file to\\messaging service};
      \draw[arrow] (request)    -- (evaluate);
      \draw[arrow] (evaluate)    -- (gencode);
      \draw[arrow] (gencode) -- (checker);
      \draw[arrow] (checker)    -- (syntaxq);
      \draw[arrow] (syntaxq) -- node[lbl,right] {Yes}  (evaluate);
      \draw[arrow] (syntaxq) -- node[lbl,right] {No} (runtopas);
      \draw[arrow] (runtopas) -- (topasq);
      \draw[arrow] (topasq.east) -- node[lbl, below right] {No} ++(1.1,0) |- (analyse.west);
      \draw[arrow] (analyse) -- (acceptq);
      \draw[arrow] (acceptq.east) -- node[lbl, above] {Yes} ++(1.1,0) |- (rootobj.west);
      \draw[arrow] (rootobj)   -- (upload);
      \draw[arrow] (upload)  -- (summary);
      \draw[arrow] (topasq.west) -- (\loopx,-9.5)
                   node[lbl,pos=0.5, above] {Yes}
                   -- (\loopx,-2.) -- (evaluate.west);
      \draw[arrow] (acceptq.south) -- ++(0,-0.6)
                   -- node[lbl, above] {No} (\loopx,-10.8)
                   -- (\loopx,-2.) -- (evaluate.west); 
    \end{tikzpicture}
    \caption{As the user prompts the \gls{llm}, it will access the resources and tools available to it via an \gls{mcp} server for generation of the \gls{topas} parameter file. \gls{ks} compares the generated energy-deposit distribution with the reference.}
    \label{fig:tool-call-diagram}
\end{figure}
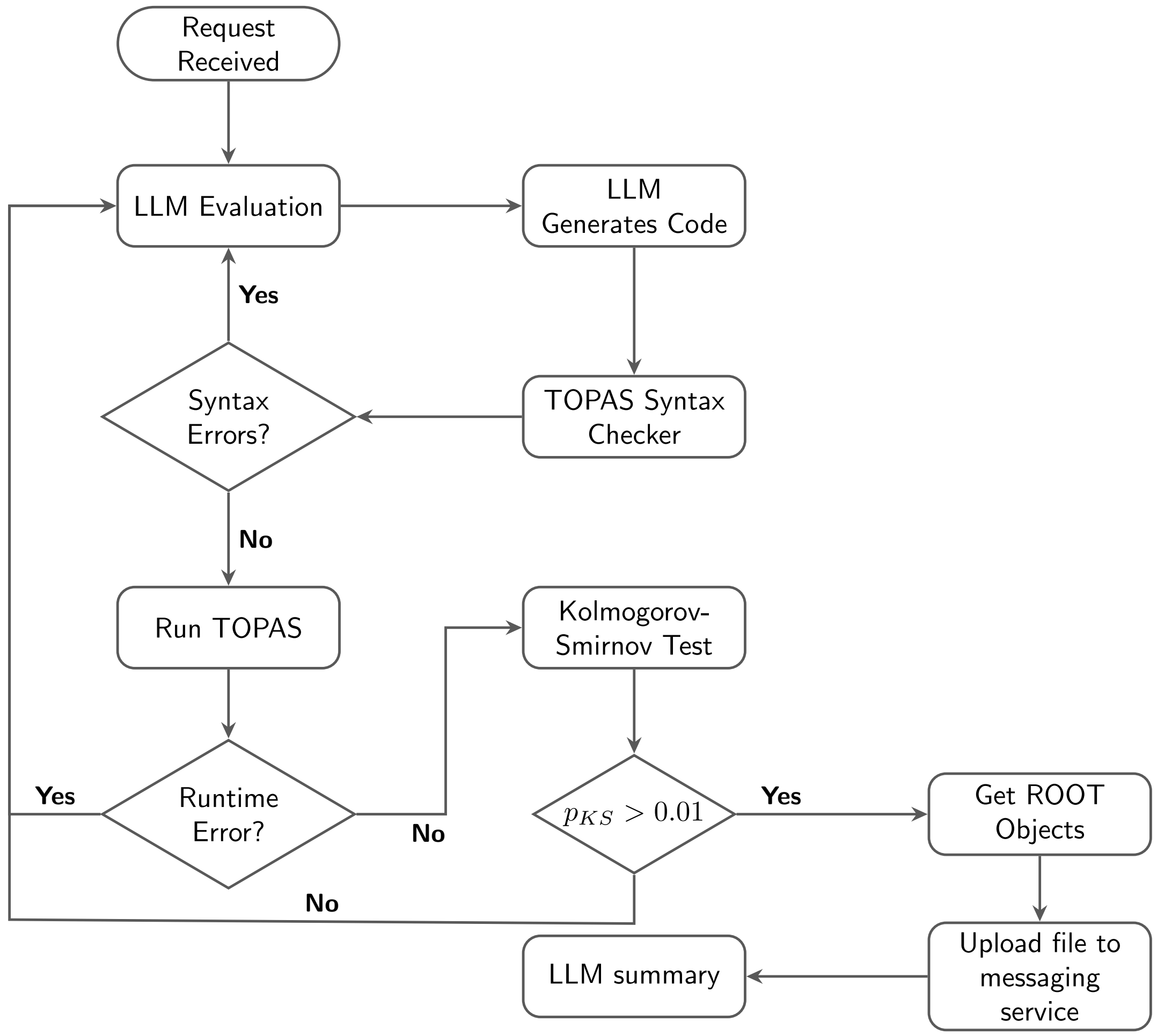

\gls{llm}s like Qwen3.6-27B are trained on a large text corpus to produce text output. The use of tools allows the model to complete tasks such as exact calculations, manipulating local files, and executing code~\cite{schick2023toolformer}. Qwen3.6-27B can generate and run \gls{mc} simulations when given tools to create and execute files in a working directory, run \gls{topas} parameter files, and read ROOT output.

Instructions are preloaded into Qwen3.6-27B's context window before it accepts a user prompt to improve the model's knowledge of detector simulations. These resources include technical documentation, examples, and rules. These also provide the model with a workflow for completing tasks and may directly instruct the model to run tools. Generating valid \gls{topas} syntax requires lists of parameter and material names, default simulation parameters, and rules for creating the files. The workflow is executed using tools, see \Cref{fig:tool-call-diagram}, in the order defined by a workflow resource, accessed through an \gls{mcp} server.

First, Qwen3.6-27B evaluates the simulation request and determines which parts of the \gls{topas} documentation are relevant. The model then generates the parameter file and verifies it with a syntax-checking tool. The syntax checker identifies mistakes in \gls{topas} parameter files, such as invalid data types, duplicate parameters, incorrectly sized vectors, missing parent components, invalid component references, incorrectly defined \gls{topas} PhaseSpace scorers, and common issues with materials, physics modules and Boolean parameters. However, it does not check whether the simulation is physically correct.

If the syntax checker returns with no errors, \gls{topas} is run with the generated parameter file. If syntax, tool, or runtime errors occur, Qwen3.6-27B re-evaluates the error and returns to the evaluation step to fix it. It then finds the energy deposit histogram within the ROOT file. A \gls{ks} test is performed to compare between model generated and the reference histograms. The generated histogram is later saved as an image, which is then uploaded to the messaging service using \gls{mcp} tools. The model then summarises its actions to the user as a final response.

\subsection{Implementation}
Prompts are received from the messaging service via its dedicated endpoint, and then passed to the \gls{llm} workflow. The workflow described in \Cref{workflow:ga} details the process of loading tools and resources, then iterating tool calls until either the prompt goal is achieved (i.e., no tool calls remain) or $N_{max} = 5$ attempts are reached. If the former condition ends the loop, then a summary of the \gls{llm}'s actions is returned to the user. An \gls{mcp} server provides the \gls{llm} access to pre-built Python tools and resources. The results of each prompt are automatically uploaded to a database to track the results of the simulation.
The data includes the model name, prompt text, response text, temperature, \texttt{top-p}, \texttt{top-k}, total context limit, response token limit, output time, output tokens, and the result of the \gls{ks} test. 

\begin{table}[H]
    \centering
    \begin{tabular}{lr}
        Parameter             & value   \\
        \hline
        temperature           & $0.15$  \\
        \texttt{num\_predict} & $15872$ \\
        \texttt{num\_ctx}     & 524288  \\
        \texttt{top-p}        & $0.3$   \\
        \texttt{top-k}        & $40$
    \end{tabular}
    \footnotesize{}
    \caption{Sampling parameter choice for running the Qwen3.6-27B.}
    \label{tab:model-sampling-parameter}
\end{table}

In the Qwen3.6-27B model file, we set the sampling parameters to the values in \Cref{tab:model-sampling-parameter}. 

Temperature~\cite{li2025temperature}: This controls how predictably the model chooses the next token. A model's unnormalised scores for each possible next token are turned into probabilities (softmax). A value above 1.0 flattens the probability distribution, and a value below 1.0 concentrates on the tokens with the highest probabilities. A higher temperature can give more varied responses; this may help with problem-solving, but it also makes nonsensical output more likely. 

Nucleus sampling, \texttt{top-p}~\cite{holtzman2020curiouscaseneuraltext}: Only the smallest set of tokens whose cumulative probability is at least p is kept. The number of candidate tokens therefore changes from step to step with the shape of the distribution.

\texttt{top-k} sampling~\cite{fan2018hierarchicalneuralstorygeneration}: Only the k most probable tokens are kept at each generation step, and the next token is sampled from them.

Context Window, \texttt{num\_ctx}: An integer giving the maximum number of tokens the model can attend to at once, counting the prompt, the conversation history, tool outputs and the response. If this is too small, model context is truncated. 

Response token limit, \texttt{num\_predict}: The maximum number of tokens the model may output in reply to a prompt. It is not the same as the total used by a single prompt, because a prompt that involves several tool calls may query the model several times. As tokens are generated sequentially, and this decoding phase dominates inference time~\cite{patel2024splitwiseefficientgenerativellm}, a lower \texttt{num\_predict} reduces the time per response and the context carried into subsequent tool-call iterations. If the limit is too small, the response, including the model’s reasoning when thinking mode is enabled, is truncated.

\begin{algorithm}[H]
\begin{algorithmic}[1]
\State \textit{Receive} user request
\State $accepted \gets$ \textbf{false}
\State $attempts \gets 0$
\State $N_{max} \gets 5$
\While{\textbf{not} $accepted$ \textbf{and} attempts $< N_{\max}$}
    \State \textit{Evaluate} request and any feedback with model
    \State \textit{Generate} \gls{topas} code with model
    \State \textit{Check} code with \gls{topas} syntax checker
    \If{Syntax errors}
        \State \textit{Return} syntax errors to model as feedback
        \State $attempts \gets attempts + 1$
        \State \textbf{continue}
    \EndIf
    \State \textit{Run} \gls{topas}
    \If{Runtime error}
        \State \textit{Return} runtime error to model as feedback
        \State $attempts \gets attempts + 1$
        \State \textbf{continue}
    \EndIf
    \State \textit{Compute} \gls{ks} test, $p_{KS}$
    \If{$p_{KS} > 0.01$}
        \State $accepted \gets$ \textbf{true}
    \Else
        \State \textit{Return} $p_{KS}$ to model as feedback
        \State $attempts \gets attempts + 1$
    \EndIf
\EndWhile
\State \textit{Get} ROOT objects
\State \textit{Upload} file to messaging service
\State \textit{Request} summary from model
\State \textbf{return} model summary
\end{algorithmic}
\caption{Simple particle detector \gls{mc} simulation with \gls{llm} }
\label{workflow:ga}
\end{algorithm}

\section{Results and discussion}
Each model-generated simulation is compared with a reference simulation of the same set-up. Every run generates $10^5$ histories with the Geant4 \texttt{g4em-standard\_opt4}  electromagnetic physics constructor. The total energy deposited per history in the generated and reference simulations is compared with a two-sample \gls{ks} test. A run is counted as successful if $p_\mathrm{KS} > 0.01$, i.e.\ if the null hypothesis that both samples are drawn from the same distribution is not rejected at the 1\% level.

A prompt is evaluated with 50 independent runs (Appendix~\ref{app:prompt1}). The generated simulation passed the \gls{ks} test in 21/50 runs (42\%, 95\% CI (29--56)\%). Intervals are Wilson score intervals. The outcome of the run is given in \Cref{tab:outcomes}.

\begin{table}[h]
\centering
\begin{tabular}{lccc}
\hline
Outcome & $e^-$ / Si \\
\hline
Passed ($p_\mathrm{KS} > 0.01$)          & 21  \\
Rejected ($p_\mathrm{KS} \le 0.01$)      & 5   \\
Simulation ran, no \gls{ks} result       & 4   \\
Context window exhausted                 & 16  \\
Unresolved syntax or runtime error       & 4   \\
\hline
Total                                    & 50  \\
Success rate (95\% CI) [\%]              & 42 (29--56) \\
\hline
\end{tabular}
\caption{Success and failure statistics over 50 runs of the electron prompt.}
\label{tab:outcomes}
\end{table}

\begin{figure}
    \centering
    \includegraphics[width=0.5\linewidth]{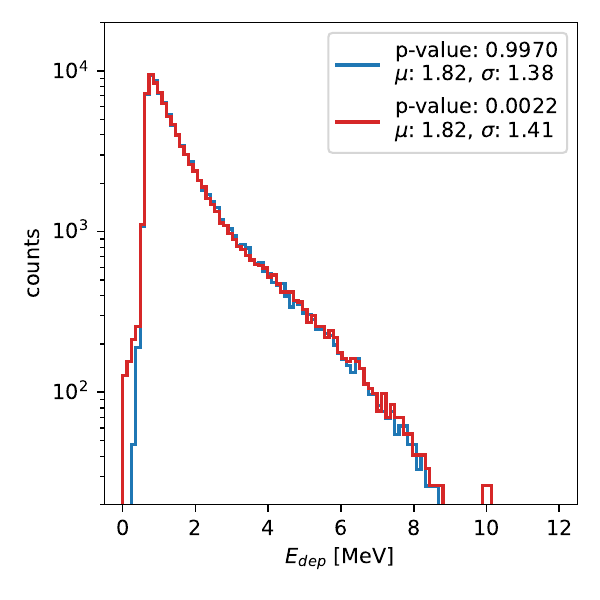}
    \caption{Accepted ($p_\mathrm{KS} = 0.997$) and a rejected ($p_\mathrm{KS} = 0.0022$) simulation generated in response to the electron prompt.}
    \label{fig:p-value-comp}
\end{figure}
\Cref{fig:p-value-comp} shows an accepted ($p_\mathrm{KS} = 0.9970$) and a rejected ($p_\mathrm{KS} = 0.0022$) simulation generated in response to the electron prompt. The two distributions have the same mean and their widths differ by about 2\%, showing that with $10^5$ histories the test is sensitive to small differences in shape. Several runs gave identical $p_\mathrm{KS}$ values; these produced identical parameter files, and with the default TOPAS random seed identical files give identical output.

The most common erroneous outcome for the prompt was exhausting the context window, which ended 16 of the 50 runs, typically after repeated documentation look-ups. The model also frequently lost track of the names of files it had generated and called a directory-listing tool to recover them, adding further tool output to the context. A further 4 runs completed the simulation but recorded no \gls{ks} result, usually because the model did not locate the reference histogram.

The median wall-clock time per run was 620\,s. 352\,s for successful runs and 691\,s for unsuccessful ones. The 50 runs took 9.0\,h of wall-clock time in total. They were served by one bot instance, which shared the A100 GPU with two other instances running the photon and proton prompts, so individual times include contention for the GPU. Successful runs ended with a median context of 88k tokens and unsuccessful runs with 182k; the 16 runs that exhausted the context window ended at 159k--219k tokens.

The prompt and its reference simulation were used both to develop the tools and resources and to evaluate them. Its success rate may therefore be optimistic for prompts not seen during development.

\section{Conclusions}
Qwen3.6-27B together with \gls{mcp} tools and access to resources on TOPAS particle detector simulation can produce simulations whose energy-deposit distribution is statistically consistent with the reference. The simulation setup has a modest success rate of 42(29--56)\% at 95\% CI for a simple prompt detailing electrons passing through a silicon detector. We hypothesise that without access to the TOPAS documentation, Qwen3.6-27B is expected to have limited knowledge of TOPAS parameter-file syntax. In our runs, syntax errors were the most common failure mode. \gls{mcp} tools and relevant documentation are intended to provide background knowledge and specialised ability. More detailed resources such as example code, further TOPAS documentation, and additional workflows for managing different requests might be required to generate more complex simulations. 

Future work would include systematic tuning of sampling parameters with larger samples. Another natural extension of this work would be testing other open-weight \gls{llm}s and comparing with closed-weight models such as Claude, GPT, and Gemini subject to cost and data-policy constraints. This work provides an initial demonstration that an open-weight model can generate simple \gls{mc} simulations, although the present study was limited by available compute and evaluation time.

\section*{Acknowledgements}
This work was supported by the EPSRC 10-week summer internship award reference EP/Z535102/1 provided for Charlie Bamber. We thank Paras Naik for a careful reading and useful comments. We also acknowledge the support of the UK Science and Technology Facilities Council (STFC).

\printbibliography
\appendix

\section{Model file and system prompt}\label{app:systemprompt}
\lstset{
    breaklines=true,
    breakatwhitespace=true,
    prebreak=,
    postbreak={},
    breakindent=0pt,
    alsoletter={.}
}

\begin{lstlisting}
FROM qwen3.6:27b
PARAMETER temperature 0.15
PARAMETER top_p 0.3
PARAMETER top_k 40
PARAMETER num_ctx 524288
PARAMETER num_predict 15872
SYSTEM """
- When discussing particles, refer to their spin, mass and charge.
- Use LaTeX for all mathematical expressions and quantum state notation.
- Think very carefully before calling functions.
- If running ROOT, run with root -b -q.
- Never use emojis
- Always tell the user explicitly the names of all files generated and their paths. Do this with any prompt and dont create any files without informing the user. 
- Refer to tools and resources when preforming simulations, never make predictions based on training data.
- Inform the user of any and all errors encoutered in either python or topas code. 
- If an errors is encountered in a generated topas parameter, write a general description of the error using Edit_Topas_Errors_File to avoid the issue in the future. 
- A simulation producing 0 histories is always an error. Do not report it as success.
- The correct parameter for number of events is: i:Ts/NumberOfHistoriesInRun
- Never use a custom source component without verifying it exists in the geometry.
- Use So/Demo as the particle source unless explicitly told otherwise.
- Always save the raw data in the form of energy, xyz positions, and time. 
- When completing simulations return a histogram of the results (of energy unless otherwise specified)
You have access to tools. You MUST follow these rules without exception:
RULE 1 - FILE CREATION IS ALWAYS A TOOL CALL:
  If the user asks you to save, write, create, or generate a file:
  - You MUST call create_file.
  - You MUST NOT output the file contents as plain text.
  - Generating content in text and then not calling create_file is a FAILURE.
  - This rule applies even if the file content is long or complex.
RULE 2 - THINK BEFORE CALLING:
  Plan the tool call sequence before acting.
RULE 3 - CHAIN TOOLS EXPLICITLY:
  Evaluate whether another tool call is required before responding.
  Produce a final text response when all tool calls for the task are complete.
RULE 4 - DECLARE TOOL USE VS INFERENCE:
  When presenting information, state explicitly whether it comes from a tool call result, or a statistical prediction based on training data.
"""
\end{lstlisting}
\section{Electron Beam and Silicon Detector}\label{app:prompt1}
The prompt describes a 10 MeV electron beam impinging on a \SI{2}{mm} thick silicon detector.

\begin{lstlisting}[mathescape=true]
Create and run a TOPAS 3.9 simulation of a 10 MeV electron beam striking a silicon detector.
Use 100 histories for an initial test. If the geometry and outputs are correct, run 100,000 histories.
Simulation setup:
    $\bullet$ Vacuum world with 5 m clearance in each direction
    $\bullet$ Silicon detector centred at the origin
    $\bullet$ Detector dimensions: 2 m $\times$ 2 m $\times$ 0.2 cm
    $\bullet$ Electron source positioned 10 cm in front of the detector along the negative z-axis
    $\bullet$ Beam directed normally toward the detector
    $\bullet$ Gaussian transverse beam position with a 0.01 cm spread
    $\bullet$ Use the g4em-standard_opt4 EM module
    $\bullet$ No graphics
    $\bullet$ For the full run set ShowHistoryCountAtInterval to 10000 to reduce the output tokens
    $\bullet$ Produce a 100-bin, 0-12 MeV histogram of total energy deposited in the detector
    $\bullet$ Use only standard TOPAS 3.9 features. Do not use unsupported custom scorers
After running:
    $\bullet$ Confirm that the simulation completed successfully
    $\bullet$ List the generated ROOT files and their exact histogram and TTree keys
    $\bullet$ Confirm that both outputs are non-empty
    $\bullet$ Compare the deposited-energy histogram with the reference histogram using a two-sample Kolmogorov-Smirnov test
    $\bullet$ Report the KS statistic and p-value
    $\bullet$ Convert the generated energy histogram to a labelled PNG and upload it to Mattermost
\end{lstlisting}

\end{document}